# Thermopower and Nernst measurements in a half-filled lowest Landau level

Xiaoxue Liu[1,2], Tingxin Li[3], Po Zhang[1,2],
L. N. Pfeiffer[4], K. W. West[4], Chi Zhang[1,2] and Rui-Rui Du[1,2,3]

[1]*International Center for Quantum Materials, School of Physics,
Peking University, Beijing 100871, P. R. China*
[2]*Collaborative Innovation Center of Quantum Matter, Beijing 100871, P. R. China*
[3]*Rice University, Houston, Texas 77251-1892, USA*
[4]*Princeton University, New Jersey 08544, USA*
\* lxxue@pku.edu.cn
\# rrd@rice.edu

**Abstract**

Motivated by recent proposal by Potter *et al*. [Phys. Rev. X **6**, 031026 (2016)] concerning possible thermoelectric signatures of Dirac composite fermions, we perform a systematic experimental study of thermoelectric transport of an ultrahigh-mobility GaAs/Al$_x$Ga$_{1-x}$As two dimensional electron system at filling factor $v = 1/2$. We demonstrate that the thermopower $S_{xx}$ and Nernst $S_{xy}$ are symmetric and anti-symmetric with respect to $B = 0$ T, respectively. The measured properties of thermopower $S_{xx}$ at $v = 1/2$ are consistent with previous experimental results. The Nernst signals $S_{xy}$ of $v = 1/2$, which have not been reported previously, are non-zero and show a power law relation with temperature in the phonon-drag dominant region. In the electron-diffusion dominant region, the Nernst signals $S_{xy}$ of $v = 1/2$ are found to be significantly smaller than the linear temperature dependent values predicted by Potter *et al.*, and decreasing with temperature faster than linear dependence.

I. Introduction

The composite fermions (CFs) paradigm has been remarkably successful in our understanding of the fractional quantum Hall effect (FQHE) [1-3]. Halperin, Lee, and Read (HLR) proposed the CFs as being one electron interacting with two Chern-Simons fluxes in a half-filled lowest Landau level (LL) [4]. At the mean field level, at $v = 1/2$, the average Chern-Simons gauge field precisely cancels the external magnetic field, so the effective magnetic field seen by CFs is zero and the electronic state of half-filling is regarded as a compressible metallic state of CFs with a distinct Fermi surface. When filling factors deviate from $v = 1/2$, CFs execute the semi-classical cyclotron orbits under an effective magnetic field $\Delta B = B - B_{1/2}$, resembling cyclotron motion of electrons under low magnetic field. Results of major experimental studies concerning $v = 1/2$ states including surface acoustic wave [5,6], magnetic focusing [7], and geometric resonance [8] are remarkably consistent with HLR theory.

Son recently emphasized [9,10] that, contrary to the belief that the particle-hole symmetry （PHS） should be preserved for the spin-polarized lowest half-filled Landau level states in the limit of negligible Landau level mixing [11-16], PHS is not explicitly built-in by HLR theory. Son proposed a particle-hole symmetric quasi-fermionic picture for $v = 1/2$ — massless Dirac composite fermions (DCF), and the Dirac nature means that there exists a Berry phase of π around the Fermi surface of CFs [9,10]. On the other hand, it has been pointed out that the proposal of DCF is not entirely different from HLR theory, since it could evolve into the picture of HLR by introducing particle-hole symmetry broken mass [9, 10]. The DCF framework has attracted much attention, largely because that it may reveal deep connections between the fractional quantum Hall effect and other physical systems, such as time reversal symmetry protected topological insulators [17-24]. From the point view of PHS, the work [9, 10] has since stimulated several proposals [25-27] to experimentally examine the evidence for PHS, or lacking of it, in a half-filled lowest Landau level.

Electrical transport has been a common and most important probe in studies concerning the $v = 1/2$ states [28-32], however, the difference in transport data between HLR and DCF theory, as proposed in [9], is difficult to discern in realistic experiments. On the other hand, as recently proposed by Potter *et al.* [25], Nernst measurement is a direct and quantitative probe for Berry phase. Note that the Nernst signal should probe the Berry phase around the CF Fermi surface, not necessarily the Dirac dispersions, so it should reveal general fermiology of composite fermions around $v = 1/2$.

We now briefly review thermoelectric transport of electrons in a temperature gradient and under a perpendicular magnetic field. Electrons will move from the hotter end to the colder end by thermally diffusion or momentum transfer with phonons (called phonon drag). Note that, among these two contributions, the electron-diffusion is dominant at the low temperature region, and the phonon drag plays a predominant role at increasing temperatures. A build-in electric field pointing to the cold end will be generated, so the electrical current can be written as $j = \hat{\sigma}(E) + \hat{\alpha}(-\nabla T)$, where $\hat{\sigma}$ and $\hat{\alpha}$ are electrical conductivity and thermal conductivity tensors, respectively. When $j = 0$, corresponding to the situation in which the diffusion current equals to the drift current but with an opposite sign, the formula above could be written as $E = \hat{S} \nabla T$, where $\hat{S} = \hat{\alpha}/\hat{\sigma}$ or $\hat{S} = \hat{\alpha} \cdot \hat{\rho}$ is the Seebeck tensor. Here $\hat{\rho}$ is the electrical resistivity tensor, $\hat{\rho} = \hat{\sigma}^{-1}$. Thermopower $S_{xx}$ refers to the diagonal part of $\hat{S}$ and Nernst-Ettingshausen $S_{xy}$ corresponds to the off-diagonal part of $\hat{S}$, where $S_{xx} = \rho_{xx}\alpha_{xx} - \rho_{yx}\alpha_{yx}$ and $S_{xy} = \rho_{xy}\alpha_{xx} + \rho_{xx}\alpha_{xy}$.

According to Potter *et al.*, HLR and DCF can be distinguished by thermoelectric measurements [25]. Specifically, the Nernst coefficient $S_{xy}$ in the diffusion - dominant region, where phonon drag is negligible, has a more direct and sensitive relationship with Berry phase around the Fermi surface of CFs. $S_{xy}$ is non-zero for the DCF theory with a PHS protected π Berry phase, and zero for PHS-broken HLR theory without the Berry phase [25]. However, a recent analysis [33] indicates that HLR theory has an emergent PHS and is equivalent to DCF theory in the limit of long-wavelengths and low-energies. When HLR theory is treated properly, Nernst coefficients of $v = 1/2$ for HLR theory should be non-zero as well [33]. Accordingly, the purpose of thermoelectric experiments should be to study the conditions under which PHS is preserved in realistic material systems, rather than to distinguish between the two theories.

In this paper, we systematically investigate the dc-transport and thermoelectric properties of the $v = 1/2$ state in a high-mobility two - dimensional electron system hosted in GaAs/Al$_x$Ga$_{1-x}$As materials. In our thermoelectric transport measurements, thermopower $S_{xx}$ and Nernst $S_{xy}$ clearly exhibit, respectively, symmetric or anti-symmetric signals with respect to $B = 0$ T, over the measured temperature ($T$) range. At the filling factor $v = 1/2$, where composite fermions form a compressible Fermi liquid in a zero effective magnetic field, thermopower $S_{xx}$ shows a linear relation with $T$ in the diffusive region, while it shows a power law $T$-dependence for the phonon-drag region. This is analogous to the thermopower $S_{xx}$ of electrons in a zero magnetic field, and consistent with previous experimental results [34-36]. As for the measured Nernst signals $S_{xy}$ of $v = 1/2$, in the phonon - drag region, they exhibit an expected power law relation with $T$. In the electron-diffusion dominated region, for the particle-hole symmetric $v = 1/2$ state, a Nernst signal $S_{xy}$ with linear $T$ dependence is predicted [25, 33]. However, the Nernst signals of $v = 1/2$ in this regime are found by our experiment to be significantly smaller than the values with linear temperature dependence predicted by Potter *et al.*, and it decreases with temperature faster than linear dependence.

## II. Experimental Method

### A. Sample Characterization

The data presented here are obtained from a high-mobility GaAs /$Al_{0.24}Ga_{0.76}As$ heterojunction wafer grown by molecular beam epitaxy. After a brief illumination from a red light-emitting diode the density and mobility (measured at 60mK) are 0.92 $\times 10^{11}$ cm$^{-2}$ and $1.1\times 10^7$ cm$^2$/V.s, respectively. A 100 μm wide Hall bar mesa was patterned on a piece of wafer (size of 10 mm $\times$ 1.7 mm $\times$ 0.5 mm) by photolithography and wet etching. Electrodes were defined by e-beam lithography, followed by Ni/GeAu evaporation and annealing process. A heater made of Ti/Au films was fabricated on one end of the sample, while the other end was indium-soldered to a copper cold finger which serves as the thermal ground. A schematic of the device is shown in Fig. 1(a). The measurement was carried out in a dilution refrigerator equipped with 8.5 T superconducting solenoid.

To characterize the samples, we performed the electrical transport measurements using the standard lock-in technique with an excitation current of 10 nA at $f = 17$ Hz. The electrical transport results are shown in Fig. 1 (b), which shows well-developed high-order FQHE states, attesting to the high mobility and the high degree of density homogeneity.

**B. The measurement method**

In thermoelectric transport experiments, thermometers are usually attached to the back of the sample, but this will induce a strain effect for high-mobility GaAs/$Al_xGa_{1-x}As$ heterostructure samples, or create a non-uniform temperature gradient. To determine temperature differences without these drawbacks, we adopted the method of integrating the thermal conductance along the sample [36]. So before the thermoelectric measurement is carried out, we first measured the temperature dependence of the sample's thermal conductivity $\kappa$, which is dominated by phonon thermal transport.

The procedure for determining thermal conductivity $\kappa$ of our sample is as follows.

First, the temperature of the cold finger is maintained at $T_i$ (which is read by a calibrated ruthenium oxide sensor), and the resistance minimum $R_{xx}$ at $v = 4/3$ is measured. Note that $R_{xx}$ at $v = 4/3$ has a strong temperature dependence over the measurement temperature range between 100 mK and 350 mK, so $R_{xx}(4/3)$ as a function of $T$ can be used as an effective thermometer. Next, $R_{xx}(4/3)$ is measured at the temperature of cold finger $T_i + \Delta T$, where the temperature increment $\Delta T < 10\% \, T_i$. The above measurement is carried out without applying any power to the heater so that the temperature of the cold finger equals the temperature of 2DEG. Subsequently, we hold the temperature of the cold finger at $T_i$ again, and adjusted the power $\dot{Q}$ applied to the heater until $R_{xx}(4/3)$ becomes close to that at temperature $T_i + \Delta T$. The temperature gradient arising from the introduction of power $\dot{Q}$ is thus determined. According to $K = \dot{Q}/\Delta T$, we calculate the thermal conductance $K$, which is between the middle of two measurement contacts and the cold end. The thermal conductance $K$ is further divided by the geometric factors of sample, to obtain the thermal conductivity $\kappa$. By repeating the above procedure at different $T_i$, the temperature dependence of thermal conductivity $\kappa$ can be systematically determined, which is shown in Fig. 2. As shown in Fig. 2, the thermal conductivity $\kappa$ is found to follow a power law against T: *i.e.*, $T^{2.6}$. The power exponent is reasonably close to that expected for phonon dominated heat transport.

To perform the thermoelectric measurement, a controllable temperature gradient $\nabla T$ was established along the Hall bar by applying a low frequency ($f = 7.3$ Hz) ac current to the heater. The thermal voltage $\triangle V$ between a pair of contacts along the gradient was measured by the lock-in technique at the frequency of $2f = 14.6$ Hz. The temperature value was obtained by integrating the thermal conductance $K$ of the sample, combining the temperature of cold finger and the applied power of the heater. The temperature or voltage gradient was further calculated according to the dimension of the Hall bar. When sweeping the perpendicular magnetic field, the thermopower $S_{xx}$ and Nernst $S_{xy}$ will then be determined by $S_{xx} = \triangledown V_{xx}/\triangledown T_x$ and $S_{xy} = \triangledown V_{xy}/\triangledown T_x$, respectively. Note here an "open circuit" condition is satisfied assuming that the lock-in amplifier has infinite input impedance.

### III. Results and Discussions

Figure 3(a) displays the thermopower $S_{xx}$ as a function of positive magnetic field $B$ at different temperatures. Meanwhile, Nernst $S_{xy}$ varying with sweeping $B$ at different temperatures are shown in Fig. 4(a) and 4(b). The thermopower $S_{xx}$ and Nernst $S_{xy}$ show, respectively, symmetric or antisymmetric patterns with respect to $B = 0$ T. Both thermopower $S_{xx}$ and Nernst $S_{xy}$ increase in magnitude with increasing temperature.

Comparing the thermopower $S_{xx}$ in Fig. 3(a) and resistivity $\rho_{xx}$ in Fig. 1(b), we observe that thermopower $S_{xx}$ exhibits oscillations of integer quantum Hall effect (IQHE) and fractional quantum Hall effect (FQHE), very much like $\rho_{xx}$. This is consistent with previous studies [34-36]. The IQHE and FQHE oscillations of $S_{xx}$ have been well explained by existing theories. In particular, the diffusion thermopower $S_{xx}^d$ of 2DEG in the QHE regime is given by entropy per particle (quasiparticle) per charge [37-40]. The proportional expression between diffusion thermopower and entropy is in accordance with that of non-interacting electrons in a zero magnetic field predicted by the Mott formula [41]. Entropy will vanish when the chemical potential $\mu$ is in the gap of incompressible states, and attain maxima once $\mu$ centered in the extended states of LLs. On the other hand, in the phonon- drag region, extended states of LLs with maximal density of states will be more likely to be scattered by phonons than localized states between LLs, leading to the oscillations of phonon-drag thermopower $S_{xx}^g$ in $B$ (similar to that of $S_{xx}^d$) [42-43]. However, although $S_{xx}^g$ and $S_{xx}^d$ have similar oscillations, their temperature dependences are entirely different: the diffusion $S_{xx}^d$, which is proportional to entropy, varies linearly with $T$, while the phonon-drag $S_{xx}^g$ shows a power law relation (with exponent ~3) as a function of $T$ [39, 41].

Figure 3(b) depicts the temperature dependence of thermopower $S_{xx}$ of $v = 1/2$, where blue circles with error bars represent low-temperature thermopower $S_{xx}$, and red ones correspond to that at higher temperatures. Note that, error bars represent the standard deviation from the average of $S_{xx}$ (shown as circles). Quite clearly, at low temperatures $T < 160$ mK, electron-diffusion is dominant, so the measured $S_{xx}$ shows a linear relation through the origin as a function of $T$. With increasing temperature, phonon-drag thermopower contributes more to the signals, leading to $S_{xx} \propto T^{2.9}$. The $T$ dependence obtained from our measurements coincides well with previous results [34-35]. This comparison also serves a test to our temperature calibration procedure. Notably, the T-dependence of thermopower $S_{xx}$ of the $v = 1/2$ compressible Fermi liquid are similar to those of non-interacting 2DEG at $B = 0$ T [39, 41,44].

According to the prediction by Cooper *et al.* [39], diffusive thermopower $S_{xx}^d$ at $v = 1/2$ can be regarded as entropy per quasiparticle per charge and can be written by the following expression, which is analogous to the Mott-formula of 2DEG at $B = 0$ T,

$$S_{xx}^d |_{v=1/2} = -\frac{\pi k_B^2 m_{CF}(1+p_{CF})}{6\hbar^2 en}T. \quad (1)$$

Here $n$ is the density of 2DEG, $m_{CF}$ is the effective mass of composite fermions determined by electron-electron interactions and $p_{CF}$ stands for the impurity scattering parameter of CFs. Due to the weak energy dependence of impurity scattering rate of CFs [39], we take $p_{CF} = 0$ here and apply Eq. 1 to our linear fit in Fig. 3(b). Note that the resulting slope (hence also the m$_{CF}$ value) of the fit may be slightly dependent on the range of data points included. We find that the m$_{CF}$ value spans roughly 0.92 $m_e$ to 0.97$m_e$ for including more data points towards the crossover point, where $m_e$ is the mass of free electrons. The Fig. 3(b) shows the fit to all 6 points in the range, yielding $m_{CF} \sim 0.97 m_e$. Overall, the uncertainty of the mass value due to fitting is within 10%. This mass value is smaller than, but roughly agrees with the previous report $m_{CF}$ ($v = 1/2$) $\approx$ 1.5$m_e$ ($p_{CF} = 0$) obtained from thermopower measurement of 2DHS [34]. It is larger than $m_{CF}$ ($v = 1/2$) $\approx$ 0.64 $m_e$ from magnetotransport measurement of 2DES around $v = 1/2$ [28]. Note that to facilitate the comparison we scale the $m_{CF}$ ($v = 1/2$) values to the same magnetic field $B = 7.6$ T by $m_{CF} \propto B^{1/2}$.

We now turn to the Nernst measurement results of $v = 1/2$ shown in Fig. 4(a) and 4(b). In realistic samples, the $S_{xx}$ and $S_{xy}$ components could couple to each other. Considering that Nernst $S_{xy}$ is antisymmetric while the thermopower component $S_{xx}$ mixed into $S_{xy}$ is symmetric with respect to $B = 0$ T, by taking the sum and difference of the raw Nernst $S_{xy}$ data under positive and negative magnetic field, we derive the "decoupled" Nernst $S_{xy}$, which is shown in Fig. 4(a) and 4(b). As shown in Fig. 4(a), for high temperature region $T > 140$ mK, the Nernst signals in the vicinity of $v = 1/2$ state increase rapidly with increasing $T$. As for the Nernst signals $S_{xy}$ below 130 mK shown in Fig. 4(b), the measured traces at different temperatures are shifted vertically for clarity.

Theoretically, in the diffusion dominant regime, the semiclassical longitudinal and transverse thermopower of 2DEG under a low perpendicular magnetic field satisfy the generalized Mott formula

$$S_{ij}^d(\varepsilon_F, T, B) = -\frac{\pi^2 k_B^2}{3e} \rho_{ik} [\frac{d\sigma}{d\varepsilon}]_{kj} |_{\varepsilon=\varepsilon_F} T \quad , \tag{2}$$

where $\varepsilon_F$ is the Fermi energy of electrons, $\rho$, $\sigma$ are the electrical resistivity and conductivity tensors, respectively [37-39,45]. Moreover, Ref. [46] showed that Eq. (2) should be valid for weakly disordered non-interacting 2DEG in the QHE regime. In the QHE regime with high magnetic field, since $\rho_{xx} \ll \rho_{xy}$ and $\alpha_{xx} \ll \alpha_{xy}$, we obtain $S_{xx}^d \approx -\rho_{xy}\alpha_{xy}$ or $S_{xx}^d \propto \rho_{xy}\frac{d\sigma_{xy}}{dB}$. Therefore, $S_{xx}^d$ will oscillate similarly to resistivity $\rho_{xx}$ as a function of magnetic field $B$. As for the diffusion Nernst $S_{xy}^d$, theories predict that $S_{xy}^d \propto \rho_{xx}.d\sigma_{xy}/dB + \rho_{xy}.d\sigma_{xx}/dB$ [37-38, 42]. Namely, when sweeping magnetic field in the QHE regime, a typical $S_{xy}^d$ trace is expected to exhibit first a positive peak, then zero at the position of the maxima of $\rho_{xx}$ or $\sigma_{xx}$, and finally a negative peak. In our experiments, as shown in Fig. 4(b), such feature of $S_{xy}^d$ can be clearly observed in the vicinity of the IQH state $v = 1$ ($B \sim 3.5$T).

Since the FQHE states around $v = 1/2$ can be viewed as the IQHE states of composite fermions, we apply Eq. (2) for analyzing thermoelectric transport in the FQHE regime around $v = 1/2$. We calculated $\rho_{xx} \cdot d\sigma_{xy}/dB + \rho_{xy} \cdot d\sigma_{xx}/dB$ by using the longitudinal resistivity $\rho_{xx}$, Hall resistivity $\rho_{xy}$ of $T = 60$ mK; the calculated trace is shown in top of Fig. 4(b). Remarkably, calculated electron diffusion Nernst signals $S_{xy}^d$ exhibits a series of oscillations in the FQHE regime around $v = 1/2$, and the period and dip positions of these oscillations (marked with blue vertical dashed lines in the figure) coincide well with our measured data. As shown in Fig. 4(b), these oscillations become more pronounced with the increasing of temperature, qualitatively consistent with the proportional relation between $S_{xy}^d$ and temperature. We found a reasonable agreement between our experimental results and theoretical predictions, confirming that the measured Nernst signals at low temperatures (<130mK) are indeed electron-diffusion dominant.

We now quantitatively analyze the data in the vicinity of $v = 1/2$. The $S_{xy}$ versus $B$ traces at the five lowest temperatures in the field range between 7 T and 8.3 T are plotted in Fig. 5(a); only the two lowest temperature traces are shifted vertically. We note that here the measured Nernst signals $S_{xy}^d$ at $v = 1/2$ drops precipitously towards low $T$. Around 100 mK the Nernst signals are of the same order of magnitudes as the background fluctuations (the amplitude of background voltage fluctuations is within $\pm 5$ nV).

We plot the Nernst signals of $v = 1/2$ against $T$ in Fig. 5(b). The higher $T$ data (red solid circles) show a power law relation $T^{3.2}$ with $T$ down to a temperature $T \sim$ 140 mK. At lower $T$ (in the electron diffusion regime), the predicted Nernst signals should crossover to a $T$-linear relation, as sketched by a dashed line in Fig. 5(b). Indeed the data (blue solid circles) begin to deviate from the $T^{3.2}$ fitting line at $T = 130$ mK. However, remarkably, the data eventually become significantly smaller than the predicted linear values (the blue dashed line), or for that matter even smaller than the values extrapolated from $T^{3.2}$.

At this point we do not have a concrete explanation for the observation of $S_{xy}^d$ in the electron - diffusion regime. A nearly vanishing $S_{xy}^d$ could indicate that the PHS is broken for $v = 1/2$ state in our 2DEG system. A possible mechanism for PHS-broken is inter-Landau-level mixing. Due to Coulomb interactions in the 2DEG system, inter-Landau -level mixing should exist in a finite magnetic field. For our low density 2DEG sample, $v = 1/2$ state is at $B = 7.6$ T, which corresponds to a substantial level of Landau level mixing. On the other hand, in GaAs/AlGaAs 2DEG systems, the phonon-drag to diffusive crossover occurs at a very low temperature (below 140mK), where the electron - diffusion dominated Nernst signals are rather small, with their magnitudes comparable to the measurement systematic errors in our experiments.

## IV. Conclusions

In summary, we have measured the longitudinal thermopower $S_{xx}$ and transverse Nernst $S_{xy}$ of compressible Fermi liquid states at filling factors $v = 1/2$ in a high-mobility GaAs/Al$_x$Ga$_{1-x}$As two-dimensional electron system. In the experimental temperature range, the thermopower $S_{xx}$ and Nernst $S_{xy}$ as a function of $B$ show respectively the expected even or odd symmetry with respect to $B = 0$ T.

Thermopower $S_{xx}$ of $v = 1/2$ present a power law relation with $T$ in the phonon-drag dominant region, and a linear dependence on $T$ in the diffusion dominant region, which is consistent with previous studies [34-36]. Furthermore, given the linear $T$ fit of the diffusion thermopower and the generalized Mott formula [41], we determine the effective mass of composite fermions of $v = 1/2$ to be $0.97m_e$ for $v = 1/2$ at $B = 7.6$ T, where $m_e$ is the mass of free electrons.

As for the Nernst signals $S_{xy}$ of $v = 1/2$, in the phonon-drag dominant region, the Nernst signals $S_{xy}$ are non-zero and have a power law dependence on $T$. In the electron - diffusion dominant region, the $S_{xy}$ show a series of oscillations in the FQHE regime around $v = 1/2$, which is consistent with the calculated results based on the generalized Mott formula. However, the measured diffusive Nernst signal of $v = 1/2$ is much smaller than the $T$-linear values predicted by relevant theory [25], and decrease with temperature faster than the linear dependence.

In this experiment the $v = 1/2$ state was set at a modestly-high magnetic field $B = 7.6$ T, so the Landau level mixing is not negligible. The influence of Landau level mixing to PHS - broken is an interesting theoretical issue which should be further addressed. We believe that the present study can provide useful guide for more refined experiments. To increase the electron diffusion Nernst signals, the $S_{xy}^d$ of $v = 1/2$ may be measured in Si/SiGe heterostructures. Unlike the piezoelectric GaAs/AlGaAs systems, the phonon drag in the Si/SiGe heterostructures is suppressed significantly below 1 K [47-49]. As a result, the $S_{xy}^d$ of $v = 1/2$ could be obtained at higher temperatures with correspondingly larger diffusion Nernst signals.

**V. Acknowledgements**

We acknowledge valuable discussions with A. H. MacDonald and J. K. Jain. We thank Changli Yang and Jian Mi for technical discussions. The work at Peking University was funded by NBRPC (No. 2014CB920901). The work at Rice was funded by NSF Grant No. DMR-1508644 and Welch Foundation Grant No. C-1682. The work at Princeton University was funded by the Gordon and Betty Moore Foundation through the EPiQS initiative Grant No. GBMF4420, by the National Science Foundation MRSEC Grant No. DMR- 1420541, and by the Keck Foundation.

Feng, and J. C. Maan, *Thermopower of a p-type Si/Si$_{1-x}$Ge$_x$ heterostructure*, Phys. Rev. B. **69**, 195306 (2004).

**Figure Captions**

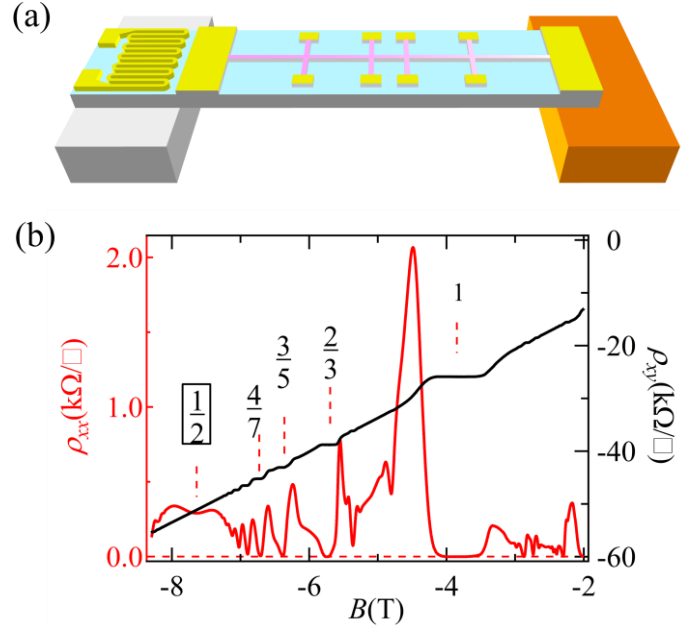

**FIG.1.** (a) Sketch map of the device used in our thermoelectric transport measurement. (b) Longitudinal resistivity $\rho_{xx}$ and Hall resistivity $\rho_{xy}$ vs magnetic field at $T = 60$ mK.

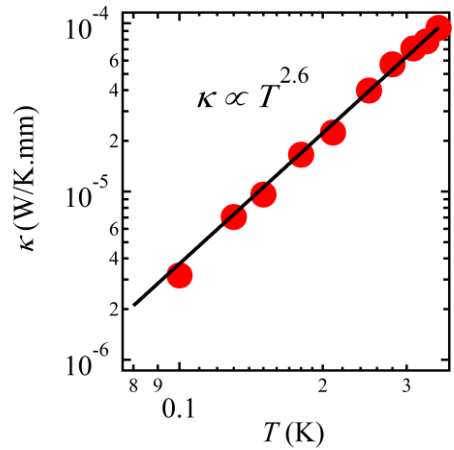

**FIG. 2.** The thermal conductivity $\kappa$ shown as red circles as a function of temperature. The black line is the fit through origin to the data and exhibits a power law relation of $T^{2.6}$.

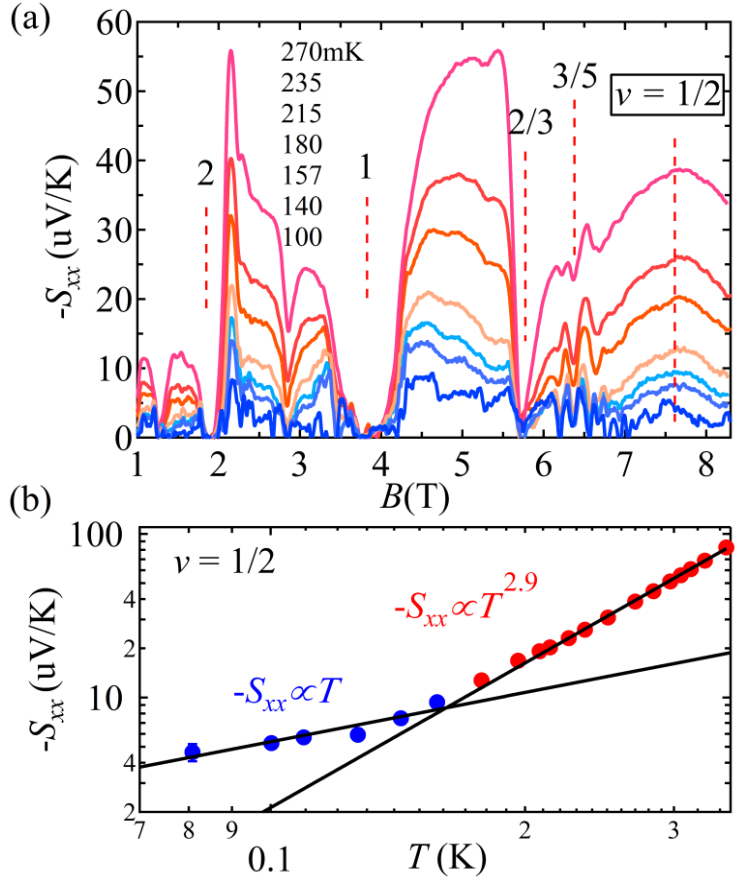

**FIG. 3.** (a) Longitudinal thermopower $S_{xx}$ vs magnetic field $B$ at different temperatures. Temperature labeled corresponds to traces from top to bottom. Several IQHE and FQHE states are marked. (b) The temperature dependence of $S_{xx}$ at $v = 1/2$. Two black lines are the fit through origin to the data in diffusion (blue circles) and phonon drag (red circles) dominant regions, respectively. The $S_{xx}$ of $v = 1/2$ shows a linear $T$ dependence in diffusion dominant region, while shows a relation of $T^{2.9}$ for phonon drag dominant region. Note that the fitting to the linear regime is in lower temperature.

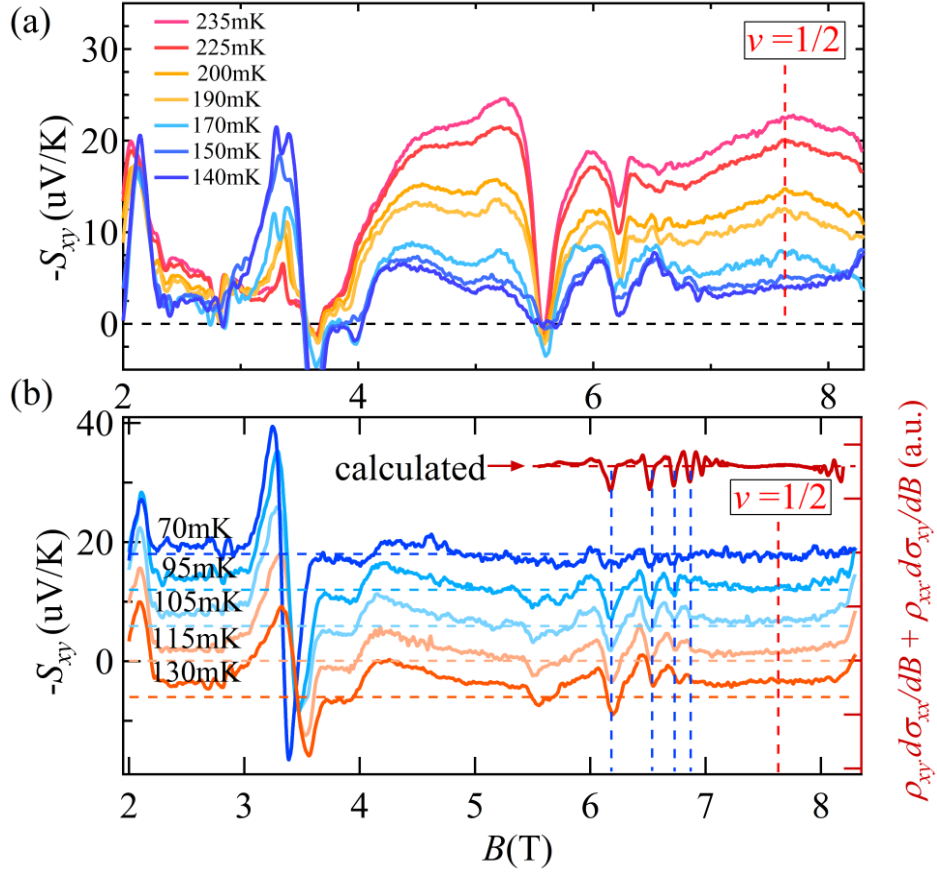

**FIG. 4.** (a) Nernst signals $S_{xy}$ vs magnetic field $B$ at different temperatures. Temperatures labeled correspond to traces from top to bottom. The $v = 1/2$ state is also labeled. (b) The topmost trace shows the calculated $\rho_{xx} \cdot d\sigma_{xy}/dB + \rho_{xy} \cdot d\sigma_{xx}/dB$ vs $B$ in the FQHE regime around $v = 1/2$. The measured Nernst $S_{xy}$ vs $B$ below 130 mK are shifted vertically for clarity. Their corresponding base lines are shown as horizontal dashed lines and have the same color with measured traces. The $v = 1/2$ state and oscillations in the FQHE regime around $v = 1/2$ state are marked by vertical dashed lines.

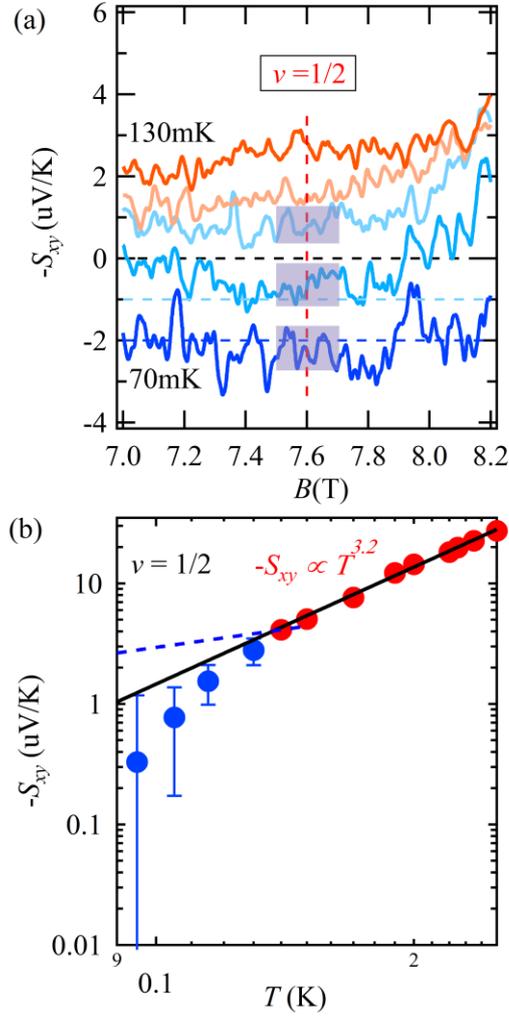

**FIG. 5.** (a) This shows a detail of Fig. 4(b), which is in the vicinity of $v = 1/2$. The $v = 1/2$ state is labeled. Only the two lowest temperature traces are shifted vertically. Their corresponding base lines are shown as horizontal dashed lines and have the same color with measured traces. Around $v = 1/2$, the three lowest temperatures traces are covered by transparent rectangle. The rectangular width describes the maximum amplitude of signal fluctuations around $v = 1/2$ (between 7.5T and 7.7T). (b) The temperature dependence of Nernst $S_{xy}$ at $v = 1/2$. The data below 130mK are shown as blue circles with error bars. The error bar here marks the uncertainty range described by rectangular width shown in Fig. 5(a). The blue dashed line shows the predicted linear temperature dependent diffusion Nernst signals. The black line is the fit through origin to data above 140mK shown as red circles, which shows a power law relation

of $T^{3.2}$ with $T$.